\documentclass[%
 aip,
 amsmath,amssymb,
 reprint,%
]{revtex4-1}

\usepackage{graphicx}
\usepackage{dcolumn}
\usepackage{bm}
\usepackage{xcolor}

\usepackage[utf8]{inputenc}
\usepackage[T1]{fontenc}
\usepackage{mathptmx}

\begin{document}

\preprint{AIP/123-QED}

\title[Oblique Shock Breakout from Uniform Density]{Oblique Shock Breakout from a Uniform Density Medium}

\author{Itai Linial}
    \email{itai.linial@mail.huji.ac.il}
\author{Re'em Sari}%
 
\affiliation{Racah Institute of Physics, The Hebrew University, Jerusalem 91904, Israel}%

\date{\today}

\begin{abstract}
The emergence of a shock from a medium with a free surface is an important process in various astrophysical phenomena. It generates the first light associated with explosions like supernovae and Gamma Ray Bursts. Most previous works considered planar or spherical geometries, where the shock front is parallel to the surface, and emerges simultaneously from all points. Here we study the hydrodynamics of an oblique planar shock breaking out from the planar surface of a uniform density ideal gas with adiabatic index $\gamma$. We obtain an analytic solution to the flow as a function of the angle between the plane of the shock and the surface $\beta$.
We find steady state solutions (in a frame moving with the intersection point of the shock and the surface) up to some critical angle ($\beta_{max}=63.4$ degress for $\gamma=5/3$ and $\beta_{max}=69.3$ degrees for $\gamma=4/3$).
We show how this analytic solution can be used in more complicated geometries where the shock is not planar, giving the exact profile of the outermost breakout ejecta. We apply our analytical results to a few realistic problems, such as underwater explosions, detonation under the surface of an asteroid, or off center detonations in a uniform sphere.\end{abstract}

\maketitle

\section{Introduction}
As a strong shock wave travelling in a medium reaches its free surface, matter swept by the shock is ejected into the vacuum. If the shock is radiation dominated, this process is accompanied by a flash of photons, escaping outwards to the vacuum as the shock approaches the surface. Shock breakout is thus an important ingredient in various astrophysical phenomena, as it generates the earliest light associated with cataclysmic cosmic events. For example, in type-II supernovae, the first light signal is produced by the breakout of the supernova shock from the stellar surface \citep{Colgate:1974,Falk:1978,Chevalier:1976,Matzner:1999,Soderberg:2008,Mazzali:2008,Modjaz:2009,Nakar:2010}. 

During shock breakout, fast moving material is ejected to the surrounding region. This ejecta may interact with the tenuous circumstellar material, and result in the emission of early time afterglow radiation associated with some types of stellar explosions \citep{Chevalier:1982,Chevalier:1994,Matzner:1999,Soderberg:2008,Mazzali:2008,Modjaz:2009}.

The topic has been extensively studied over the past decades, in the Newtonian regime \citep{Colgate:1974,Falk:1978,Chevalier:1976,Matzner:1999,Nakar:2010} and more recently, in the relativistic regime \citep{Nakar:2012}. These works, however, assume that the shock is strictly parallel to the surface, treating the problem as one-dimensional. In reality, some degree of obliquity will always be present. \cite{Matzner:2013} consider the breakout of an oblique shock from a medium with a power-law density. They discuss the implications of obliquity on the shock emission in such scenarios.

In this paper, we discuss the effects of obliquity on the breakout of an adiabatic, non-relativistic shock from a medium of uniform density. Most astrophysical scenarios have a declining density profile towards the edge of the medium, and therefore we leave discussion on the radiative properties of such shocks to a future paper. First we consider an infinite oblique shock with no curvature, i.e. a shock with a planar front, emerging from a planar surface. Working in a frame that follows the intersection of the shock with the boundary, we derive an exact steady-state hydrodynamic solution to the two-dimensional flow around the intersection. Such solutions exists if the angle, $\beta$ between the shock wave and the surface is below some critical value $\beta_{max}$.

We then discuss few examples where either the shock inside the medium is curved or that it is emerging from a curved surface, or both. We show that while we no longer have an analytic solution to this more general problem, the vicinity of the shock-surface intersection point is accurately described by our analytic solution. We are therefore able to provide exact analytical results to the envelope of the ejected material in this general case.

The paper is organized as follows: in section \ref{sec:analytical_solution} we analytically derive the hydrodynamic solution describing the flow due to an oblique breakout of planar shock from a planar surface. In section \ref{sec:Applications} we discuss applications of the steady-state solution to a few realistic problems. We conclude by discussing our results in section \ref{sec:discussion}.

\section{Analytic Steady-State Solution:
A Planar Shock Emerging Obliquely from A Planar Surface} \label{sec:analytical_solution}
\subsection{Analytical derivation}
Consider a cold, inviscid material of uniform density $\rho_0$, filling the upper half-space, $y>0$. An oblique planar shock wave, propagating towards the planar boundary surface $y=0$, intersects it along a line parallel to the $z$-axis. Since the system is translationally invariant along the $z$ direction, the problem is two-dimensional, as shown in figure \ref{fig:schematic_oblique_shock}. We denote the angle between the shock and the surface by $0 < \beta \leq \pi/2$.
If the shock propagates at velocity $v_{sh}$ normal to the shock, the point of intersection of the shock and the $x$-axis moves at the pattern speed $v_0 = v_{sh}/\sin{\beta}$.

The flow around the intersection point is steady in a frame moving at velocity $v_0 \mathbf{\hat{x}}$, hereafter, the steady-state frame. In addition, the lack of a natural length scale implies that the flow depends solely on the polar angle $\theta$, measured from the $x$-axis, with the origin at the intersection point of the shock and the surface. Thus, in the steady-state frame, the flow is described by the velocity components, $v_r$ and $v_\theta$ and by the density $\rho$, all of which are functions of $\theta$ alone. Assuming a polytropic equation of state, $P \propto \rho^\gamma$, the continuity and momentum equations reduce to \textcolor{black}{(see appendix \ref{sec:appendix})}
\begin{equation} \label{eq:continuity_uniform_density}
    \frac{\partial v_\theta}{\partial \theta} = -v_r - v_\theta \frac{1}{\rho} \frac{\partial \rho}{\partial \theta} \,,
\end{equation}
\begin{equation} \label{eq:momentum_r_uniform_density}
    \frac{\partial v_r}{\partial \theta} = v_\theta \,,
\end{equation}
\begin{equation} \label{eq:momentum_theta_uniform_density}
    v_\theta \left( \frac{\partial v_\theta}{\partial \theta} + v_r \right) + \frac{1}{\gamma} \left[ \frac{1}{\rho} \frac{\partial \rho}{\partial \theta} c_s^2 + 2 c_s \frac{\partial c_s}{\partial \theta} \right] = 0 \,,
\end{equation}
where $c_s$ is the speed of sound, $c_s^2 = \partial P / \partial \rho$. Assuming the flow is adiabatic, the Bernoulli equation implies constant enthalpy along streamlines
\begin{equation}
    c_s^2 + \frac{\gamma - 1}{2} \left( v_r^2 + v_\theta^2 \right) = const \,.
\end{equation}
The cold incoming flow has velocity $v_0$ in the negative $x$ direction, and thus the speed of sound at any position is given by
\begin{equation} \label{eq:c_s_bernoulli}
    c_s^2 = \frac{\gamma - 1}{2} \left( v_0^2 - v_r^2 - v_\theta^2 \right) \,.
\end{equation}

We differentiate equation \ref{eq:c_s_bernoulli} with respect to $\theta$ and substitute equations \ref{eq:continuity_uniform_density} and \ref{eq:momentum_r_uniform_density} in equation \ref{eq:momentum_theta_uniform_density}, to obtain the following simple relation
\begin{equation} \label{eq:uniform_density_result}
    (c_s^2 - v_\theta^2) \frac{\partial \rho}{\partial \theta} = 0 \,.
\end{equation}
Equation \ref{eq:uniform_density_result} is satisfied either by demanding $\rho = const.$, or by setting $v_\theta = \pm c_s$. As a consequence, whenever the density is non-uniform, the flow must be supersonic, as $|\mathbf{v}|^2 = v_r^2 + v_\theta^2 \geq v_\theta^2 = c_s^2$. Following equations \ref{eq:continuity_uniform_density} and \ref{eq:momentum_r_uniform_density}, when the density is constant, the velocity vector is constant, i.e., material flows along straight streamlines
(note that this does not mean $v_r$ and $v_\theta$ are constant along streamlines). This is clear, since constant density implies constant pressure, and hence there is no acceleration.

\subsection{Boundary conditions}
In the steady-state frame, the cold incoming flow intersects the stationary shock at angle $\beta$. Just before the shock, at angle $\theta=\beta_-$,
the density is $\rho(\beta_-)=\rho_0$, and the velocity components are $v_r(\beta_-) = -v_0 \cos{\beta}$ and $v_\theta(\beta_-)= v_0 \sin{\beta}$. In the immediate region past the shock, the flow is compressed to density $\rho(\beta_+)=\rho_0 (\gamma+1)/(\gamma-1)$, the velocity component parallel to shock is unchanged, $v_r(\beta_+)=v_r(\beta_-)$ and the tangential velocity component is reduced to $v_\theta(\beta_+) = v_\theta(\beta_-) (\gamma-1)/(\gamma+1)$. Hence, right after the shock, the flow is deflected below the horizontal direction, forming an angle $\alpha_p$ with the negative $x$-axis
\begin{equation} \label{eq:alpha_p_deflection}
    \alpha_p = \beta - \arctan{ \left( \left( \frac{\gamma-1}{\gamma+1} \right) \tan{\beta} \right)} \,.
\end{equation}
Finally, the flow velocity and the speed of sound past the shock are given by
\begin{equation} \label{eq:v_p}
    v_p = v_0 \sqrt{1-\frac{4\gamma}{(\gamma+1)^2} \sin^2{\beta}} \,,
\end{equation}
\begin{equation} \label{eq:c_sp}
    c_{sp} = \frac{\sqrt{2\gamma (\gamma-1)}}{\gamma+1} v_0 \sin{\beta} \,.
\end{equation}
The values at the shock set the boundary conditions for the flow equations \ref{eq:continuity_uniform_density}-\ref{eq:momentum_theta_uniform_density}.

\subsection{Flow structure}

\begin{figure}[!ht]
    \includegraphics[width=1.04\columnwidth]{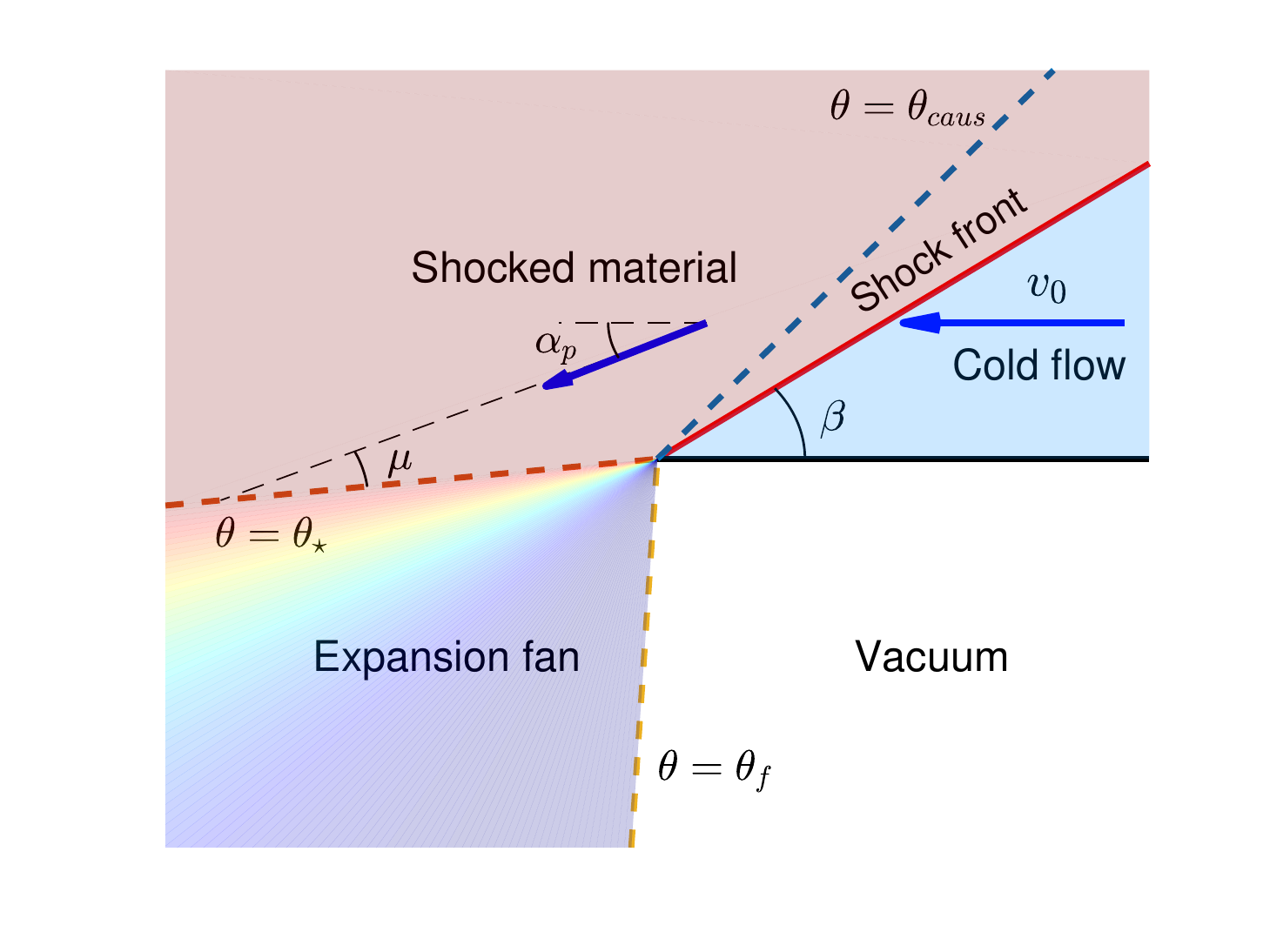}
    \caption{Schematic structure of the flow around the intersection of an oblique shock and a boundary. We work in the steady-state \textcolor{black}{frame}, moving with pattern velocity $v_0$ in which the shock-boundary intersection point is stationary. Cold material flows towards the shock front, where it is heated and deflected by an angle $\alpha_p$. The shocked material follows straight streamlines, up to angle $\theta_\star$, at which the flow begins to rarefy in a \textcolor{black}{Prandtl-Meyer} expansion fan. The region between the shock and $\theta_{caus}$ is causally connected with the shock - sound waves emitted in this angular region reach the shock. The expansion fan terminates at angle $\theta_f$, at which the flow is cold and purely radial, \textcolor{black}{having velocity $v_0$.}}
    \label{fig:schematic_oblique_shock}
\end{figure}

We now turn to identify regions of uniform density, $\partial \rho / \partial \theta = 0$, and regions at which $v_\theta = c_s$, where the density varies, satisfying equation \ref{eq:uniform_density_result}. The flow's density upstream of the shock is constant, $\rho = \rho_0$, and streamlines are straight, parallel to the $x$-axis. Material in the immediate region past the shock has uniform density and velocity, \textcolor{black}{with the flow deflected by an angle $\alpha_p$}. However, while the velocity vector is fixed in that region, its radial and tangential components vary. \textcolor{black}{Immediately past the shock, $v_\theta < c_{sp}$, and $v_r < 0$. However, as $\theta$ increases, $v_\theta$ also increases up to an angle $\theta = \pi/2 + \alpha_p$, at which the flow is purely tangential, with $v_\theta = v_p$. At even higher angles, $v_\theta$ decreases, and the radial velocity component is positive and increasing.} Eventually, $v_\theta = c_s$, and at this angle, denoted as $\theta_\star$, the flow changes its character from uniform to an expansion with decreasing density. This rarefaction is nothing but a Prandtl-Meyer expansion fan \citep{Prandtl:1907,Meyer:1908}. This critical angle $\theta_\star$ could be understood as the angle at which the presence of vacuum in the lower half plane affects the flow. As we shall demonstrate, the expansion fan terminates at some angle $\theta_f$, at which the density vanishes, and the velocity is purely radial. The flow is thus divided into three angular regions, schematically depicted in figure \ref{fig:schematic_oblique_shock} and summarized in the following:
\begin{equation}
    \begin{split}
        0 \leq \theta < \beta & \qquad \frac{\rho}{\rho_0} = 1 \,, \qquad \textrm{straight streamlines} \\
        \beta < \theta \leq \theta_\star & \qquad \frac{\rho}{\rho_0} = \left( \frac{\gamma + 1}{\gamma - 1} \right) \,, \; \textrm{straight streamlines} \\
        \theta_\star \leq \theta \leq \theta_f & \qquad v_\theta = c_s \,, \qquad \textrm{expansion fan} \\
    \end{split}
\end{equation}

What is the angle $\theta_\star$ at which the expansion fan begins? 
Past the shock, matter initially flows with velocity $v_p$ along straight streamlines, directed at an angle $\alpha_p$ below the horizontal direction (equations \ref{eq:alpha_p_deflection}-\ref{eq:v_p}). The tangential velocity component $v_\theta$ \textcolor{black}{changes} with $\theta$ until $v_\theta(\theta_\star) = c_{sp}$, where the expansion fan begins, satisfying equation \ref{eq:uniform_density_result} by maintaining $v_\theta = c_s$.

\begin{figure}[!ht]
    \includegraphics[width=1.04\columnwidth]{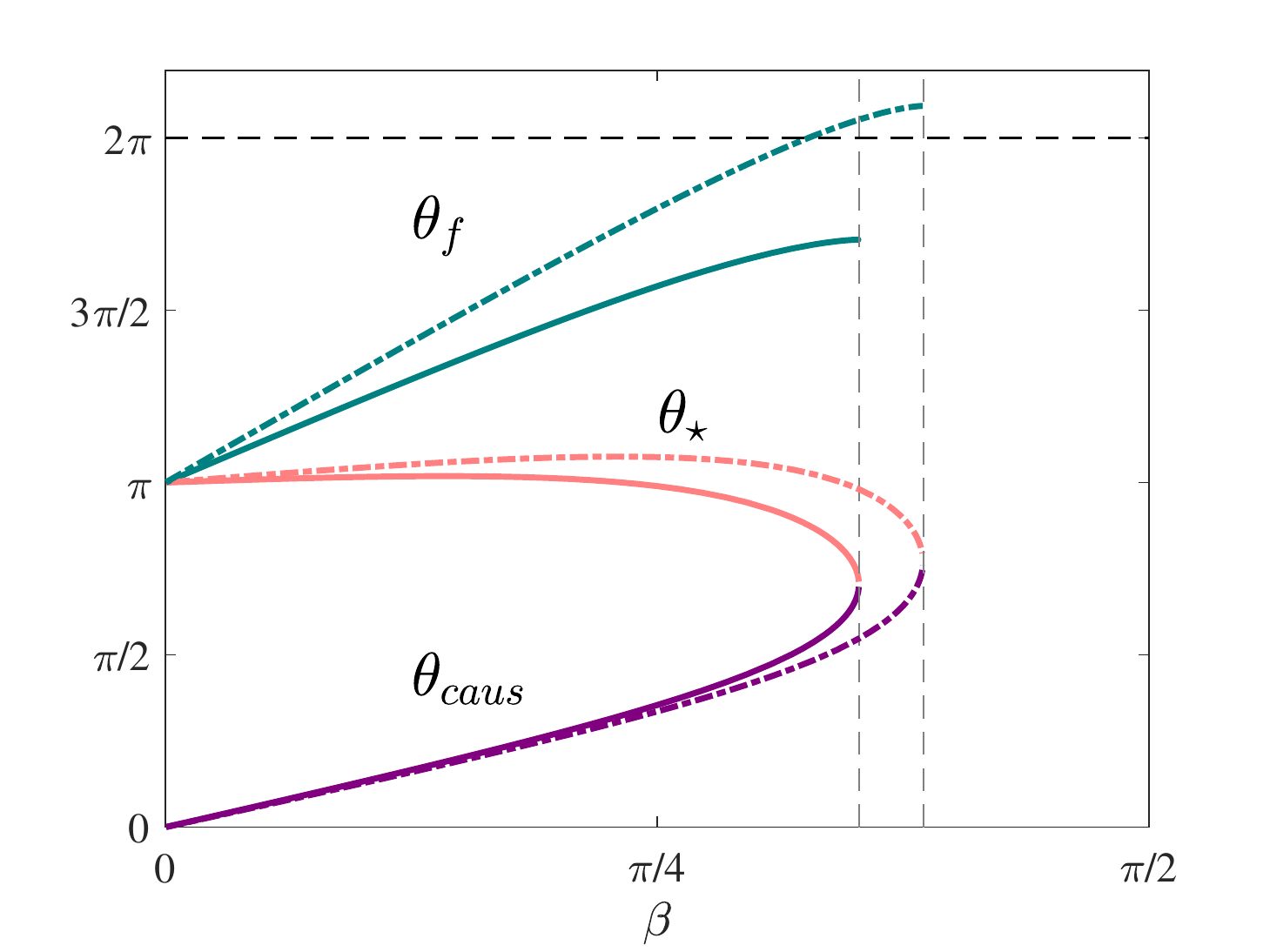}
    \caption{\textcolor{black}{Angular regions in the steady-state solution, as a function of the shock angle $\beta$. \textit{Red} is $\theta_\star$, the expansion fan's initial angle, \textit{green} is $\theta_f$, the expansion fan's final angle, and \textit{purple} is $\theta_{caus}$, the region causally connected to the shock. \textit{Solid lines} correspond to $\gamma = 5/3$ and \textit{dashed-dot} correspond to $\gamma = 4/3$.}
    Steady solutions exist for $\beta < \beta_{max}$, given in equation \ref{eq:beta_max}, at which the curves terminate (\textit{gray dashed vertical lines}). Note that for $\gamma = 5/3$, $\theta_f < 2\pi$ up to $\beta_{max}$, while for $\gamma = 4/3$, the fan expands beyond $2\pi$ for $1.026 < \beta < \beta_{max}(\gamma=4/3)$, making this range un-physical. Note that as $\beta$ approaches $\beta_{max}$, $\theta_\star = \theta_{caus}$, implying that information from the rarefied flow can reach the shock front.}
    \label{fig:Theta_star_Theta_f}
\end{figure}

The angle $\theta_\star$ can also be obtained by analyzing the propagation of sound waves emitted from the origin - the shock-boundary intersection point. The flow begins its expansion when the presence of a boundary is transmitted by sound waves. As the post-shock flow is supersonic (in the steady-state frame), sound waves emitted from the shocked boundary cannot propagate into the entire domain. Rather, a line emerging from the origin at angle $\theta_\star$, marking the transition to the expansion fan, intersects the streamlines at the Mach angle, $\mu = \arcsin{(c_s/v_p)}$, as illustrated in figure \ref{fig:schematic_oblique_shock}. 
Both descriptions lead to the same result, $\theta_\star = \pi + \alpha_p - \mu$. The angle $\theta_\star$ is therefore determined by the shock intersection angle $\beta$, as shown in figure \ref{fig:Theta_star_Theta_f}.

Solving for the expansion fan region we set $v_\theta = c_s$ in equation \ref{eq:momentum_theta_uniform_density}, to obtain the following
\begin{equation} \label{eq:density_v_theta_uniform_density}
    \frac{1}{\rho} \frac{\partial \rho}{\partial \theta} = \left( \frac{2}{\gamma - 1} \right) \frac{1}{v_\theta} \frac{\partial v_\theta}{\partial \theta} \,.
\end{equation}

Substituting equation \ref{eq:density_v_theta_uniform_density} into equation \ref{eq:continuity_uniform_density} we find
\begin{equation}
    \frac{\partial v_\theta}{\partial \theta} = -\left( \frac{\gamma-1}{\gamma+1} \right) v_r \,,
\end{equation}
which, along with equation \ref{eq:momentum_r_uniform_density} yields a set of two coupled differential equations that can be solved analytically. The boundary conditions at $\theta=\theta_\star$ are by construction
\begin{equation} \label{eq:bc_v_r_theta_star}
    v_r(\theta_\star) = \sqrt{v_p^2 - c_{sp}^2} = v_0 \sqrt{ 1 - \frac{2\gamma}{\gamma + 1} \sin^2{\beta}} \,,
\end{equation}
\begin{equation} \label{eq:bc_v_theta_theta_star}
    v_\theta(\theta_\star) = c_{sp}\, .
\end{equation}

Solving the differential equations, we find that for $\theta > \theta_\star$
\begin{equation}
    v_r(\theta) = v_0 \cos{\left(\sqrt{\frac{\gamma-1}{\gamma+1}}(\theta_f - \theta) \right)} \,,
\end{equation}
\begin{equation}
    v_\theta(\theta) = v_0 \sqrt{\frac{\gamma-1}{\gamma+1}} \sin{\left(\sqrt{\frac{\gamma-1}{\gamma+1}}(\theta_f - \theta) \right)} \,,
\end{equation}
where $\theta_f$ is the final fan angle, determined by the boundary conditions \ref{eq:bc_v_r_theta_star} and \ref{eq:bc_v_theta_theta_star}. At the end of the expansion fan, $\theta = \theta_f$ the flow is cold and purely radial, with $v_r = v_0$, and $v_\theta = c_s = 0$. The density as a function of $\theta$ is finally found by integrating equation \ref{eq:density_v_theta_uniform_density}.

\begin{figure}[!ht]
    \includegraphics[width=1.04\columnwidth]{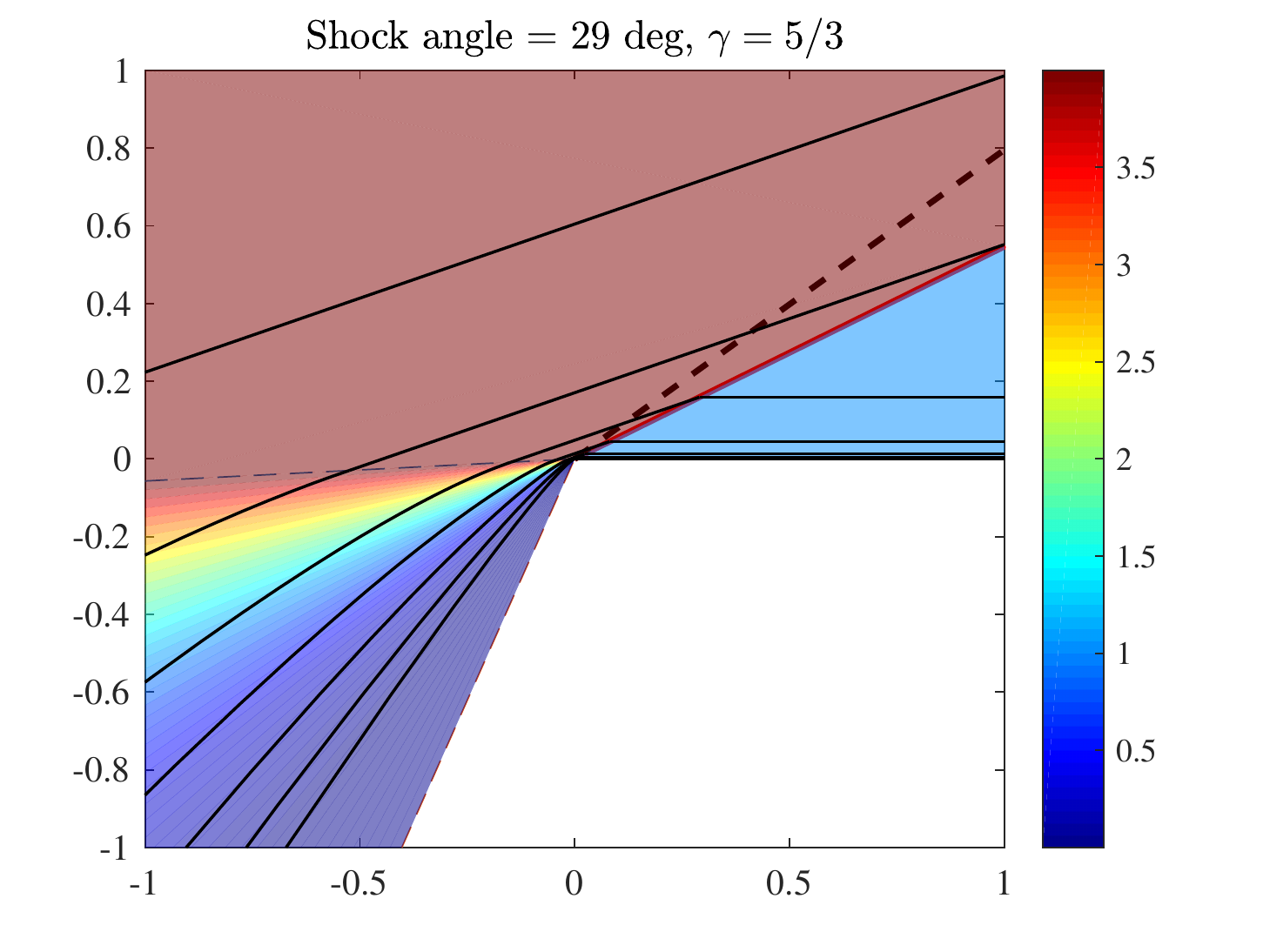}
    \caption{\textcolor{black}{The steady flow around the intersection of an oblique shock and the surface, for a shock angle $\beta=0.5 \, \rm rad$, and $\gamma=5/3$. Color indicates density, normalized by the density of the upstream material. The \textit{gray dashed line} mark the beginning of the expansion fan, $\theta_\star$ and the flow termination angle, $\theta_f$. \textit{Solid black lines} are streamlines. The streamlines are initially horizontal in the upstream and turn abruptly at the shock. Between the shock and $\theta_\star$ material flows along straight streamlines, that later bend within the expansion fan, where the flow rarefies. The fan terminates at $\theta_f$, at which the density and tangential velocity both vanish. The region between the shock and the \textit{black thick dashed line} at angle $\theta_{caus}$ is causally connected to the shock front.}}
    \label{fig:uniform_density_streamlines_small_beta}
\end{figure}

Written explicitly, the expansion fan initial and final angle are given by
\begin{multline}
    \theta_\star = \pi + \beta - \arctan{ \left( \left( \frac{\gamma-1}{\gamma+1} \right) \tan{\beta} \right)} - \\ \arcsin{ \left( \sqrt{ \frac{2\gamma (\gamma-1)}{(\gamma+1)^2 - 4\gamma \sin^2{\beta}}} \sin{\beta} \right)} \,,
\end{multline}
\begin{equation} \label{eq:theta_f}
    \theta_f = \theta_\star + \sqrt{\frac{\gamma+1}{\gamma-1}} \arcsin{\left( \sqrt{\frac{2\gamma}{\gamma+1}} \sin{\beta} \right)} \,,
\end{equation}
and are plotted in figure \ref{fig:Theta_star_Theta_f}. Figure \ref{fig:uniform_density_streamlines_small_beta} shows the flow structure for $\beta = 1/2$ and $\gamma = 5/3$, demonstrating the three flow regions: pre- and post- shock, as well as the expansion fan.

\subsection{Solution validity range}

\begin{figure}[!ht]
    \includegraphics[width=1.04\columnwidth]{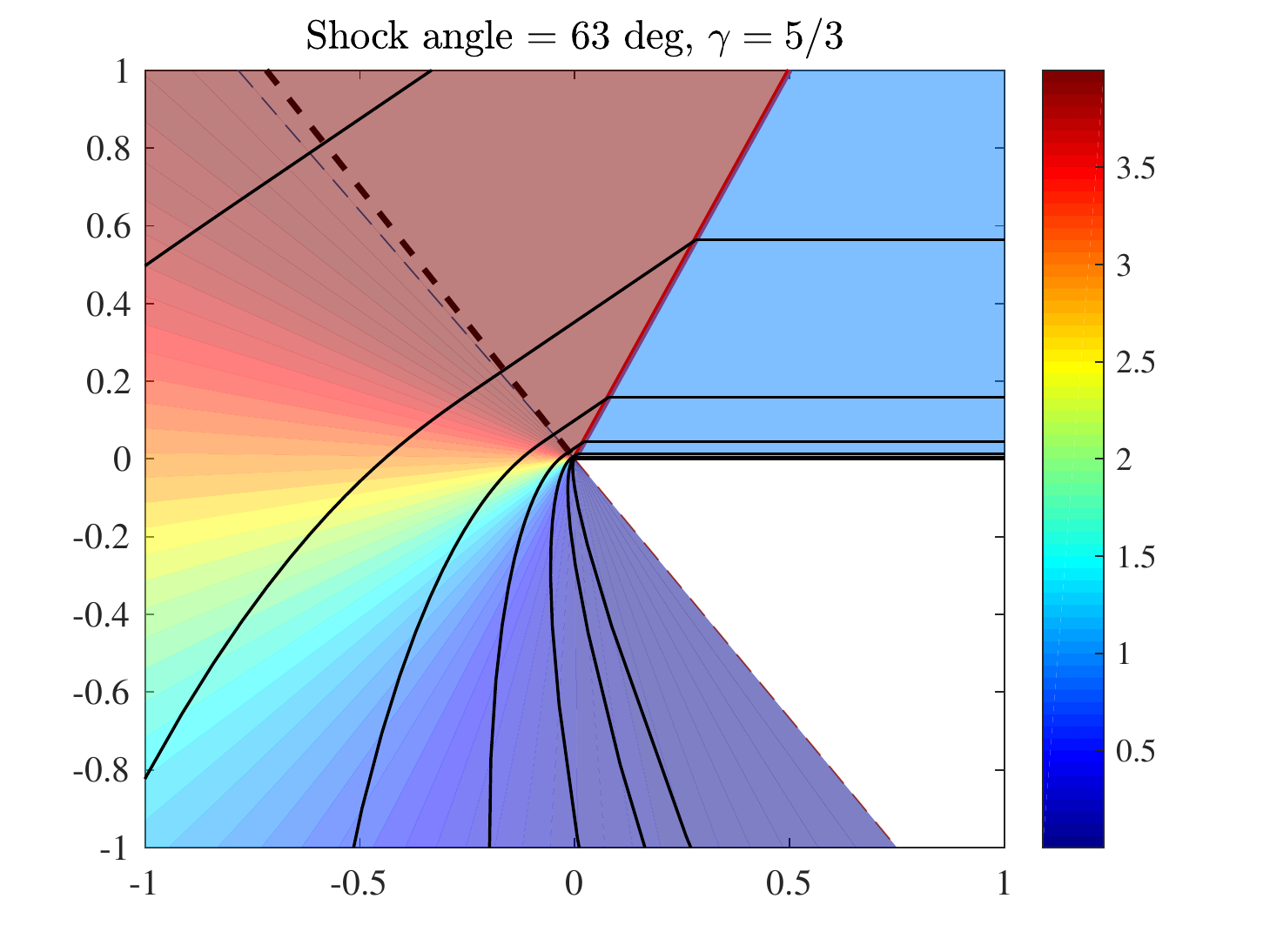}
    \caption{Same as figure \ref{fig:uniform_density_streamlines_small_beta}, just before the maximal shock angle $\beta=\beta_{max}$, defined in equation \ref{eq:beta_max}. At this angle, the Mach number immediately past the shock is $1$, and therefore the \textit{dashed line} where the expansion fan begins intersects the streamlines at $90$ degrees. The \textit{black dashed line} is $\theta_{caus}$, the extent of the region causally connected with the shock. Notice how at this angle, $\theta_\star$ is almost tangent to $\theta_{caus}$.}
    \label{fig:uniform_density_streamlines_beta_max}
\end{figure}

In order to comply with the condition $v_\theta = c_s$ at the expansion fan, a steady solution exists only for sufficiently small $\beta$, for which the post-shock flow is supersonic. The limiting angle, at which the post-shock Mach number is unity is given by solving $v_p=c_{sp}$ using equations (\ref{eq:v_p}) and (\ref{eq:c_sp}):
\begin{equation} \label{eq:beta_max}
    \beta_{max} = \arcsin{ \sqrt{\frac{\gamma + 1}{2 \gamma}} } \,.
\end{equation}
For common adiabatic indices, $\gamma = 5/3$, $\beta_{max} = 63.4$ degrees, and for $\gamma = 4/3$, $\beta_{max} = 69.3$ degrees. Figure \ref{fig:uniform_density_streamlines_beta_max} demonstrates the flow structure for $\gamma = 5/3$, when the shock intersection angle is approaching the limiting angle $\beta_{max}$. 

\begin{figure}[!ht]
    \includegraphics[width=1.1\columnwidth]{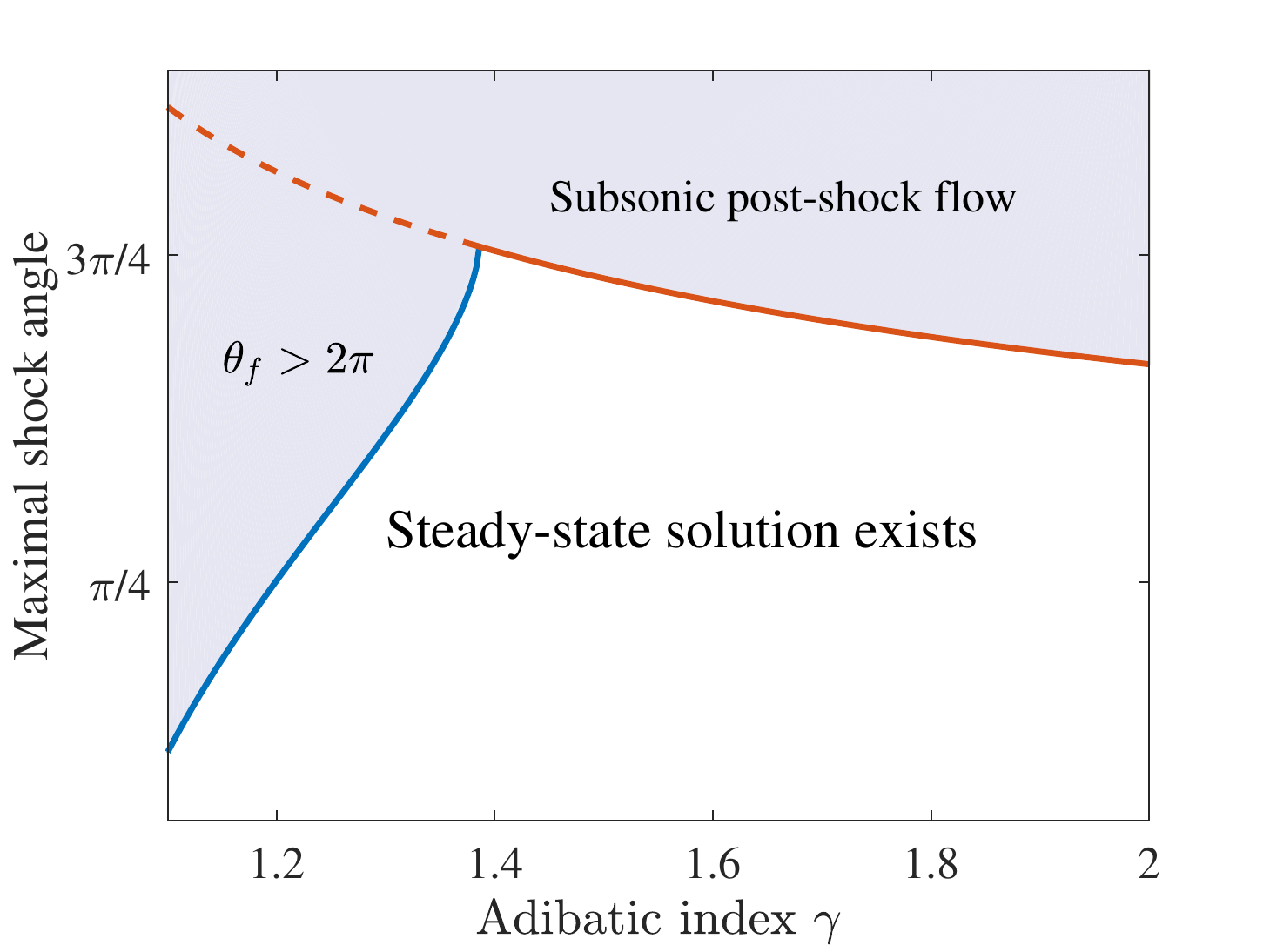}
    \caption{A steady-state flow in the vicinity of the shock-boundary intersection region exists up to some maximal intersection angle $\beta$. \textit{Shaded} regions are forbidden. When the adiabatic index is small, $\gamma < 1.386$, the limiting $\beta$ is obtained by demanding that the expanding material does not collide with the upstream region, i.e., $\theta_f < 2\pi$ (\textit{blue} curve). For $\gamma > 1.386$, a steady-state solution exists up to $\beta_{max} = \arcsin{\sqrt{(\gamma+1)/2\gamma}}$, set by the condition that the shocked material is supersonic (\textit{red} curve).}
    \label{fig:UniformDensityBetaValidity}
\end{figure}

We note that for $\gamma < 1.386$ the expansion fan extends all the way to $2\pi$ at some $\beta_{\theta_f = 2\pi} < \beta_{max}$. In such cases, the expansion fan interacts with the upstream material, acting as a precursor to the shock itself. Since our derivation assumed a cold upstream, our solution does not account for intersection angles larger than $\beta_{\theta_f = 2\pi}$. \textcolor{black}{The interaction of the fan with the upstream will result in a surface compression shock. The flow direction of the incoming material is opposite to that of the fan material along the surface, which may result in a Kelvin-Helmholtz instability due to the velocity shear. These effects have been explored in \citep{Salbi:2014}, where the emergence of oblique shocks from a medium of varying density is studied. The steady state solutions we obtain may provide some insight regarding the flow structure even past $\beta_{\theta_f=2\pi}$, as the flow is expected to change mostly near the surface.} \textcolor{black}{Nonetheless, our steady-state solution is only} valid in the range $0 < \beta < \min \{ \beta_{max} , \beta_{\theta_f = 2\pi} \}$, depending on the value of $\gamma$, as shown in figure \ref{fig:UniformDensityBetaValidity}. 

\subsection{Boundary-shock causal connection}
What makes the limiting angle $\beta_{max}$ special? Why is there no steady solution for larger intersection angles?

The downstream of a regular, non-oblique, planar shock is subsonic at the shock frame, and hence the entire downstream is causally connected with the shock front. On the contrary, for an oblique shock, only parts of the downstream can communicate with the shock. This occurs since the pattern speed $v_0$ can be significantly larger than the shock speed $v_{sh}$.

In the steady-state frame, the downstream is supersonic for small $\beta$. Hence, sound waves emitted from an arbitrary point in the downstream would extend out to a Mach cone with its axis parallel to the streamlines in the shocked region. Therefore, the region causally connected to the shock is limited to an angle $\theta_{caus} = \alpha_p + \mu$, where $\alpha_p$ is the flow deflection angle (equation \ref{eq:alpha_p_deflection}), and $\mu$ is the Mach angle. As long as the flow in the causally connected region is unaffected by the boundary, the shock front is unaware of the presence of the boundary.

The flow is first modified by the presence of a boundary at angle $\theta_\star$, at which it begins to rarefy. Thus, the angular gap between the expansion fan and the causallity region is $\theta_\star - \theta_{caus} = \pi - 2\mu$. As $\beta$ increases, the Mach number decreases, and the gap between $\theta_{caus}$ and $\theta_\star$ decreases. At $\beta = \beta_{max}$, $\mu = \pi/2$, and $\theta_{caus} = \theta_\star$. At this stage, sound waves emitted from the affected flow can marginally make it to the shock front. This behavior is demonstrated in figure \ref{fig:Theta_star_Theta_f}, where $\theta_\star$ and $\theta_{caus}$ converge to the same value at $\beta = \beta_{max}$.

\section{Applications} \label{sec:Applications}
In this section we investigate a few scenarios in which a strong shock encounters a free surface obliquely, and apply the analytic results derived in section \ref{sec:analytical_solution}.
Unlike the synthetic example of the previous section, where both the shock and the boundaries are planar, we investigate here a more general case. However, over short timescales and short distances (i.e. shorter than the radius of curvature of the surface or the shock), the solution near the the intersection of the shock wave with the boundary should follow our analytic solution. It is this intersection point from which the fastest material is ejected into the vaccuum. Since the flow then continues ballistically, it will accurately describe the external contour of the ejected material even in the general case.

We begin by the discussing the breakout to vacuum of a steady-state bow-shock in a cold medium. In the second part of this section, we apply our solution to a strong explosion occurring nearby a free surface. Finally, we consider a strong explosion within a uniform sphere, occuring off-center. Despite being a time-dependant problem, our steady-state analytical solution can be still applied to obtain some interesting predictions.

\subsection{Bow-shock breakout}
Bow shocks appear when a supersonic flow encounters an obstacle. When the flow is of finite Mach number $M$, the shock front far from the obstacle is a cone with an opening angle $\arcsin{(1/M)}$, known as the Mach cone. However, when the material is cold, i.e., $M=\infty$, the shock front forms a parabolic figure of rotation \citep{Yalinewich:2016}. For an obstacle of size $R$, the asymptotic shock shape is approximately
\begin{equation}
    x/R = a (r/R)^2\,,
\end{equation}
where the flow is directed along the positive $x$ direction, $r$ is the cylindrical radius coordinate, and $a \approx 0.53$, valid in the limit $r \gg R$. The obstacle itself is located at the stand-off distance, $z=bR$, where $b$ is an order unity constant.

If the medium is bounded by a planar surface, parallel to the direction of motion of the obstacle, the bow shock intersects the boundary and breaks out into the vacuum. As the shock is parabolic, the shock-boundary intersection angle depends on the distance of the obstacle from the surface. The further the obstacle is from the boundary, the less oblique the breakout becomes, with a smaller intersection angle $\beta$.

\begin{figure}[!ht]
    \includegraphics[width=1.1\columnwidth]{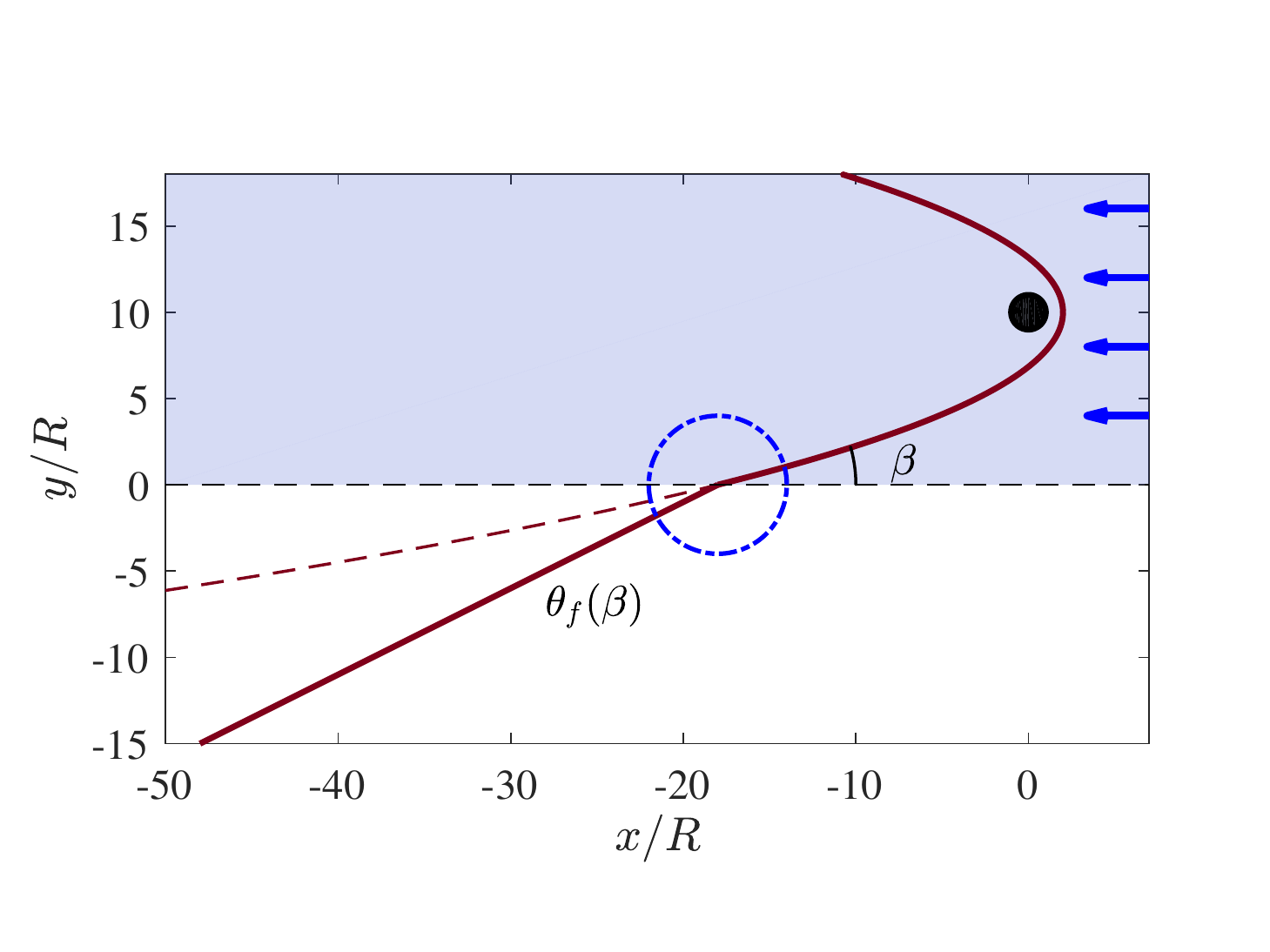}
    \caption{Illustration of a bow-shock breakout. An object of radius $R$ is placed at a distance $d=10 \, R$ from the surface of a cold material that flows past the object. A parabolic bow shock is formed around the object, intersecting the surface at angle $\beta$. This angle is independent of the existence of a surface and is dictated by the shape of the shock that would form in an infinite medium.
    A spray of material is formed around the intersection point, with its boundary forming a straight line at angle $\theta_f(\beta)$, given by equation \ref{eq:theta_f} (\textit{purple solid line} at the lower half of the figure). \textcolor{black}{Although the shock is curved, sufficiently close to the intersection point (\textit{blue dash-dotted} circle) it can be approximated as a part of a planar shock. The dynamics in this region are described by the self-similar solution described in section \ref{sec:analytical_solution}}.}
    \label{fig:BowShockGraphics}
\end{figure}

Consider a cold flow past an object of size $R$. The obstacle is placed at $y=d$, where $d \gg R$, and the material's upstream velocity is $-v_0 \mathbf{\hat{x}}$. The incoming material is bounded at $y=0$, similar to the settings described in section \ref{sec:analytical_solution}. The shock-boundary intersection occurs at $x_0 \approx -a d^2 / R$, and the intersection angle scales as $\beta \sim R/d$ (figure \ref{fig:BowShockGraphics}).

In a small region around the intersection point, the steady flow can be described locally by the analytical solution found in section \ref{sec:analytical_solution}. The flow thus forms an expansion fan towards the vacuum, terminating at an angle $\theta_f(\beta)$. Since the flow at the end of the expansion fan is cold and purely radial, the deflected material continues to propagate ballistically along a straight trajectory forming an angle $\theta_f-\pi$ with the $x$ axis, where $\theta_f$ is given by equation \ref{eq:theta_f}.

\subsection{Strong point explosion near a surface} \label{subsec:surface_explosion}
When a large amount of energy is instantaneously deposited in a small volume within a cold uniform medium, a strong shock propagates spherically away from the origin. This famous problem was originally solved by Taylor, von-Neumann and Sedov, using a self-similarity argument \citep{Taylor:1950,von_Neumann:1958,Sedov:1959}. Conservation of energy gives the scaling of the shock radius with time, and the self-similar pressure, density and velocity profiles within the shocked region can be found analytically. 

If the surrounding cold medium is bounded by a planar surface, the spherical shock breaks out obliquely into the vacuum. We use the analytical steady-state solution obtained in section \ref{sec:analytical_solution} to describe the evolution of the system after the initial shock breakout. This problem naturally arises when a strong explosion occurs near the surface of a non-stratified medium, e.g., a point explosion triggered inside the crust of an asteroid, or an underwater detonation. Note that gravity is being neglected in what follows, as we assume that the system evolves on a short time-scale, before gravity begins to play an important role.

Consider a cold material of uniform density, filling the upper half-space, $y>0$. A point explosion positioned at $(x=0,y=R_0)$ is triggered at time $t=0$. A shock wave initially expands spherically, following the Sedov-Taylor (ST) solution, until it reaches the boundary $y=0$ at time $t_0$.

We begin by briefly describing our results qualitatively. Shortly after the initial breakout at time $t_0$, the shock continues to follow the ST solution \textcolor{black}{within the original medium} at $y>0$, unaffected by the \textcolor{black}{presence of} vacuum. This stage terminates at time $t_{max}$, when information about the presence of the boundary is first delivered to the shock front.

\begin{figure}[!ht]
    \includegraphics[width=1.1\columnwidth]{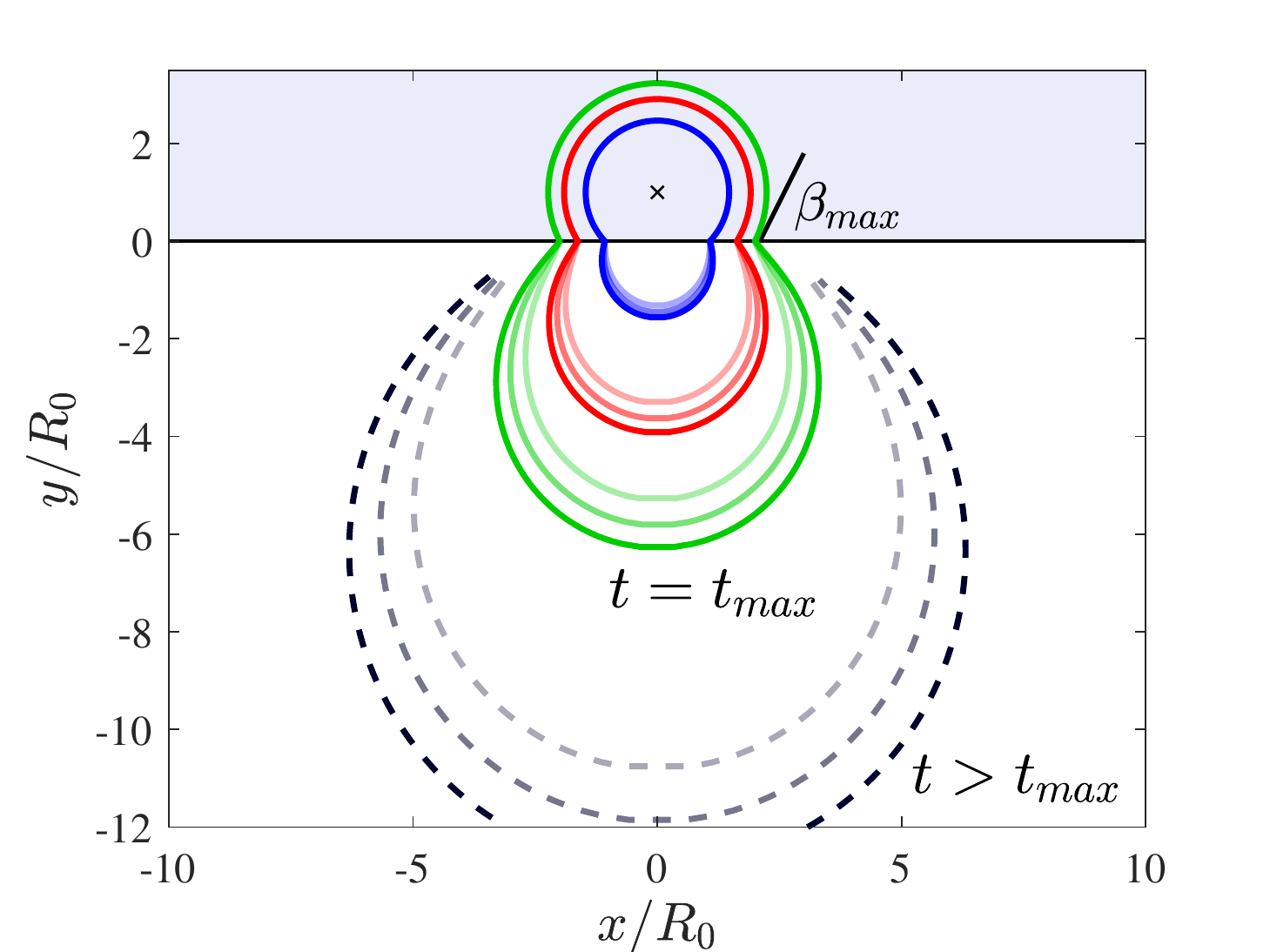}
    \caption{\textcolor{black}{An explosion is detonated in $(x=0,y=R_0)$ at time $t=0$ in a region of uniform density and $\gamma = 5/3$ occupying the half space $y>0$. Different colors correspond to different times after the explosion. Curves in the upper half-plane (with the uniform density) show the explosion's shock front, and at the lower half-plane (vacuum), we show the extent of the explosion's ejecta. \textit{Solid green} contours correspond to time $t_{max}$, and the \textit{black-gray dashed} contours correspond to time $t=1.9 \, t_{max}$. Shades of a certain color demonstrate the extent of different densities (bottom to top) - $\rho = 0$, $2\times10^{-3}$ and $2\times10^{-2}$, where $\rho = 1$ is the medium's original density. At time $t_{max}$ the shock forms an angle $\beta_{max}$ with the boundary. Our solution is valid up to this stage, and the contours at later times only describe part of the ejecta's envelope, without the region adjacent to the surface at these late times. We do not solve the shock's shape within the medium at times later than $t_{max}$.}}
    \label{fig:SphericalExplosion}
\end{figure}

Throughout this stage, $t_0 < t < t_{max}$, shocked material is ejected into the vacuum. The outermost ejecta expands and extends farther than where a spherical shock in uniform medium would have reached. The shock front and outermost ejecta are plotted in figure \ref{fig:SphericalExplosion}. At the end of this period, $t=t_{max}$, the angle between the spherical shock and the free surface equals the maximal angle, $\beta_{max}$, and the shock becomes subsonic \textcolor{black}{with respect to the matter behind it}.

At later times, $t \gtrsim t_{max}$, the shock propagation at $y>0$ is altered. Sound waves emitted from the shocked boundary catch up with the shock front, initially affecting the shock-boundary intersection region, and gradually progressing towards the entire shock front.

Quantitatively, the shock expands spherically up to time $t_0$, with its radius given by the ST solution
\begin{equation} \label{eq:ST_radius}
    R(t) = R_0 \left( \frac{t}{t_0} \right)^{2/5} \,.
\end{equation}

At times $t_0 < t < t_{max}$, the shock continues to expand according to equation \ref{eq:ST_radius} in the upper-half plane $y>0$. The shock front intersects the boundary at position $x_{sb}(t)$
\begin{equation} \label{eq:x_sb}
    x_{sb}(t) = R_0 \sqrt{\left( \frac{t}{t_0} \right)^{4/5} - 1 } \,,
\end{equation}
\textcolor{black}{forming an angle $\beta(t)$ with the surface}
\begin{equation} \label{eq:beta_spherical_explosion}
    \beta (t) = \arccos{\left( \frac{R_0}{R(t)} \right)} = \arccos{\left( \frac{t_0}{t} \right)^{2/5}} \,.
\end{equation}

Despite being a time dependant problem, and although the emerging shock is curved, our analytical steady-state solution for a planar shock (section \ref{sec:analytical_solution}) can still be utilized in this case. The flow around the shock-boundary intersection is not affected by the shock curvature, assuming that we concentrate on a region much smaller than $R$. Within this local region, matter sweeps across on timescales much shorter than $t$, the time on which the global solution evolves. Thus, the flow around the intersection point evolves gradually, making our steady-state, planar solution appropriate for analyzing the shock breakout in this scenario.

As a consequence, the terminal ejecta velocity can be deduced from the analytical solution. At time $t_0 < t < t_{max}$, the momentary pattern speed of the intersection point is $v_0 = \dot{R}/\sin{\beta}$. Within the local steady-state frame, material is accelerated and deflected to velocity $v_0$, moving at the terminal expansion-fan angle $\theta_f$. Transforming back to the lab frame, the outermost ejecta produced at time $t$ has velocity
\begin{equation} \label{eq:v_terminal_lab}
    \mathbf{v}_{terminal} = \frac{\dot{R}}{\sin{\beta}} \left( \left( 1+\cos{\theta_f (\beta)} \right) \, \hat{\mathbf{x}} + \sin{\theta_f(\beta)} \, \hat{\mathbf{y}} \right) \,.
\end{equation}
Any position $x$ along the boundary has a corresponding breakout time $t_{bo}(x)$, given by inverting equation \ref{eq:x_sb}. Matter originating from $x$ is ejected at time $t_{bo}(x)$ and propagates ballistically with the terminal velocity given by equation \ref{eq:v_terminal_lab}, reaching position 
\begin{equation}
    \mathbf{r}(x) = x \hat{\mathbf{x}} + ( t - t_{bo}(x) ) \, \mathbf{v}_{terminal}(x) \,,
\end{equation}
at time $t_{bo}(x) < t < t_{max}$. \textcolor{black}{We are thus able to accurately calculate the ejecta's envelope at these times, as demonstrated in figure \ref{fig:SphericalExplosion}}.

\textcolor{black}{Next, we estimate the density distribution of the explosion's fast ejecta, behind the outermost material. By following streamlines in our steady-state solution, we trace the position of material of a given density, $\tilde{\rho}$, after some arbitrary travel time, and obtain density contours behind the ejecta envelope. Note that this calculation is valid only for fast moving ejecta, that has low density, $\tilde{\rho} \ll \rho_0$, and travels almost ballistically after a short acceleration period. The calculation of density contours is demonstrated in figure \ref{fig:SphericalExplosion}, where we show low density contours behind the leading ejecta front.}

The solution is valid up to time $t_{max}$, at which the shock intersection angle approaches the limiting value $\beta_{max}$ (equation \ref{eq:beta_max}). Up to this stage, when $\beta$ is smaller than $\beta_{max}$, the shock front propagates sufficiently fast along the boundary, such that sound waves emitted from the shock-boundary intersection point cannot overtake the shock front. As time passes, $\beta$ increases to $\beta_{max}$, where sound waves emitted from the boundary catch up with the shock front, \textcolor{black}{and weaken it}. From equation \ref{eq:beta_spherical_explosion}, this stage terminates at time
\begin{equation}
    t_{max} = t_0 \left( \frac{2 \gamma}{\gamma - 1} \right)^{5/4} \,,
\end{equation}
at which the radius is $R(t_{max}) = R_0 \sqrt{2 \gamma / (\gamma - 1)}$. 

For typical values of $\gamma$, the shock propagates along the boundary unhampered for a fairly long time after the initial breakout. For example, for $\gamma = 5/3$, $t_{max} \approx 7.5 \, t_0$, at which the shock radius is roughly $2.2 \, R_0$.

\textcolor{black}{Since the speed of sound close to the explosion's origin is higher than at the shock front, it is not obvious why $t_{max}$ is the time at which causal connection between the shock and the rarefied flow is first achieved. Could certain sound wave trajectories originating from the surface overtake the shock earlier than $t_{max}$? Using Sedov's solution for the self-similar flow fields within the interior of the blastwave, we find that at times earlier than $t_{max}$ the shock is causally disconnected with the surface. We addressed this question numerically, by investigating the 2D propagation of sound waves within the interior of a Sedov-Taylor explosion (not shown in this work). Sound waves that pass through the origin (where the speed of sound diverges), expand spherically and arrive at the shock front at times later than $3 \, t_{max}$, and thus only provide a causal connection after that achieved at $t_{max}$}.

Even though our steady-state solution can be applied just up to time $t_{max}$, parts of the ejecta's envelope can be computed at later times. Since the fastest moving ejecta propagates ballistically, matter ejected up to time $t_{max}$ continue to follow its straight trajectory \textcolor{black}{see figure \ref{fig:SphericalExplosion}}.

Long after the shock breaks out, when the initial separation of the source from the boundary is small compared with the size of the influenced region, the flow asymptotically becomes scale-free. In this regime, the flow converges to the self-similar surface explosion problem described by \cite{Zeldovich_Raizer:1967} in the context of cratering on planetary surfaces. Hence, our solution acts as an intermediate step linking these two self-similar regimes, prior to the breakout, and long afterwards.

\textcolor{black}{Related problems have been studied in the past decades, in the context of explosions at the surface of the ocean \citep{Holt:1977}. In \cite{Falade_Holt:1975,Falade_Holt:1976}, the authors consider the interaction of a point explosion with the free surface of the ocean, and obtain the shape of the disturbed surface. In their work, they identify a self-similar flow structure around the shock-boundary intersection, similar to the solution we discuss in section \ref{sec:analytical_solution}, including the identification of a Prandtl-Meyer expansion fan. An important difference between this work and Ref, is that while we consider an underwater explosion, these past works treat an explosion set on the ocean's surface.}

\textcolor{black}{Our work describes the evolution after the shock's initial emergence at time $t_0$, and we use a series of steady-state solutions of increasing obliquity, to obtain the ejecta's shape up to time $t_{max}$. In the problem considered in \citep{Falade_Holt:1976}, there is no special length or time scale, and the flow is self similar at all times, (see also \citep{Zeldovich_Raizer:1967}). An important component in their solution is the use of a criterion proposed by Zaslavskii \citep{Zaslavskii:1963} for determining the flow structure near the surface. Stated in terms used in our paper, Zaslavskii conjectured that the shock forms tends to form an angle $\beta_{max}$ with the surface, corresponding to the largest angle steady-state solution we identify. Following section \ref{sec:analytical_solution}, at $\beta_{max}$, sound waves from the rarefied flow can marginally overtake the shock. If Zaslavskii's criterion is indeed correct, it may be used also in the case we consider, of a submerged explosion, in order to determine the evolution at times later than $t_{max}$, which we do not address in this work.}

\subsection{Off-center spherical explosion}
Oblique shocks are a natural outcome of non-isotropic or asymmetric explosions. In this section we consider the dynamics of a strong off-center explosion within a uniform sphere. Since the shock front in this case is not parallel to the surface of the sphere, breakout is not instantaneous, and matter is ejected aspherically to the surrounding region. We \textcolor{black}{apply the results derived in previous sections to obtain the exact ejecta envelope as a function of time}.

\begin{figure}[!ht]
    \includegraphics[width=1.1\columnwidth]{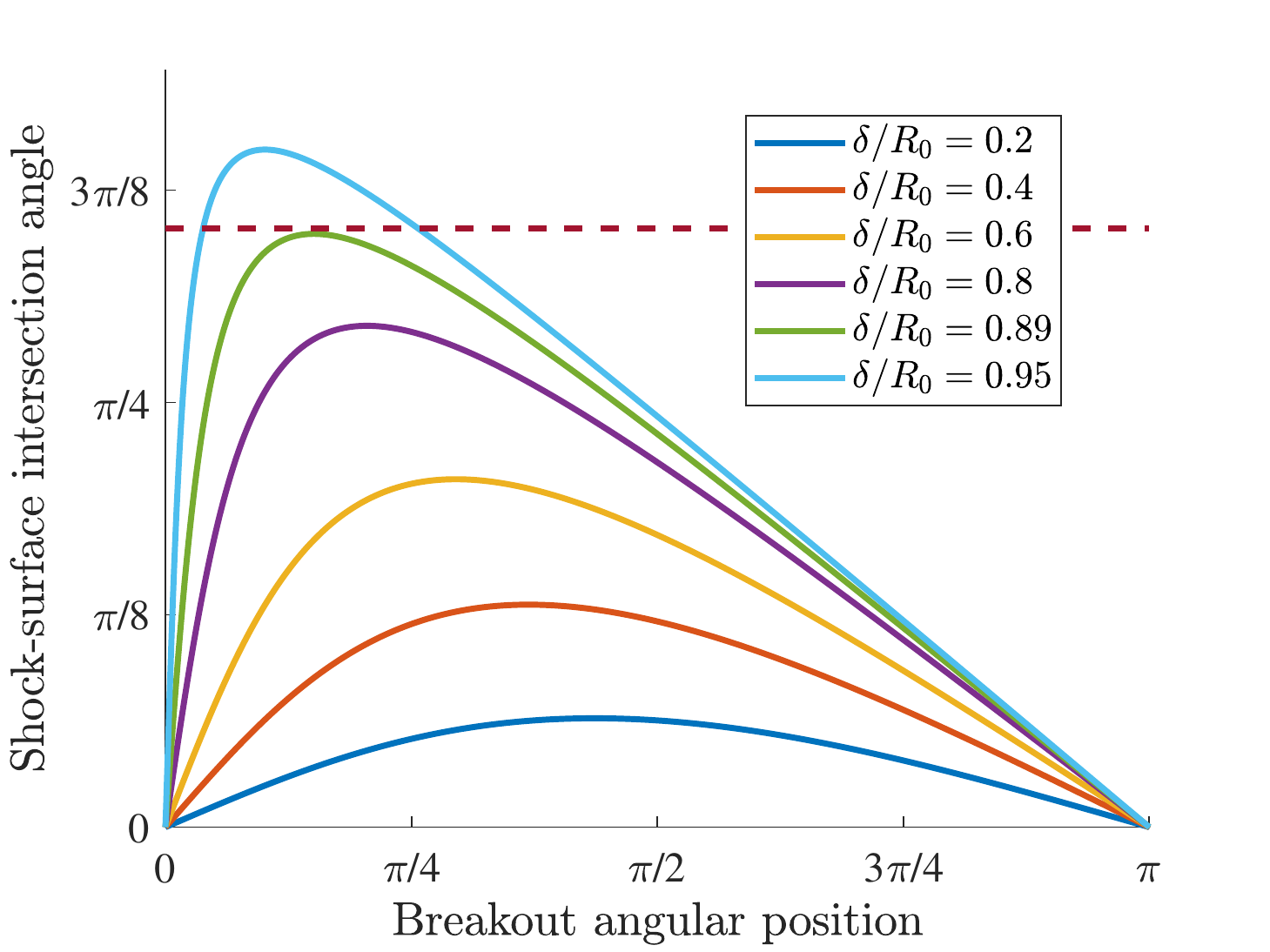}
    \caption{An explosion is detonated within a sphere of uniform density, at a distance $\delta$ from the center. The shock-surface intersection angle is plotted vs. the angular position along the surface, measured from the initial breakout point. Different \textit{solid lines} correspond to different values of $\delta$. The horizontal \textit{dashed line} is the maximal intersection angle for which the steady-state solution found in section \ref{sec:analytical_solution} exists, taking $\gamma = 5/3$. We can therefore apply this solution to calculate the ejecta's trajectory at any time, as long as $\delta < 0.89 \, R_0$.}
    \label{fig:OffsetExplosionAngleDelta}
\end{figure}

Consider a sphere of uniform density with radius $R_0$. When an off-center explosion is detonated within the sphere, the resulting shock wave breaks out obliquely to the surrounding. The shock-surface intersection angle depends on $\delta$ - the explosion's offset from the center of the sphere. The larger $\delta$ is, the larger the shock obliqueness is, as demonstrated in figure \ref{fig:OffsetExplosionAngleDelta}.

For sufficiently small $\delta$, the shock wave propagating through the sphere is a part of a Sedov-Taylor solution - the shock front is unaware of the fact that parts of the shock have already emerged from the surface. The limiting $\delta$ is found by demanding that the shock-surface intersection angle does not exceed the maximal value $\beta_{max}$ given by equation \ref{eq:beta_max}. The maximal $\delta$ is given by
\begin{equation} \label{eq:delta_max}
    \delta_{max}/R_0 = \sqrt{\frac{\gamma + 1}{2 \gamma}} \,,
\end{equation}
for $\gamma = 5/3$, $\delta_{max}/R_0 \approx 0.89$, and for $\gamma = 4/3$, $\delta_{max}/R_0 = 0.94$. The shock propagation within the sphere is unhampered even for quite large explosion offsets relative to the size of the sphere.

\begin{figure}[!ht]
    \includegraphics[width=1.1\columnwidth]{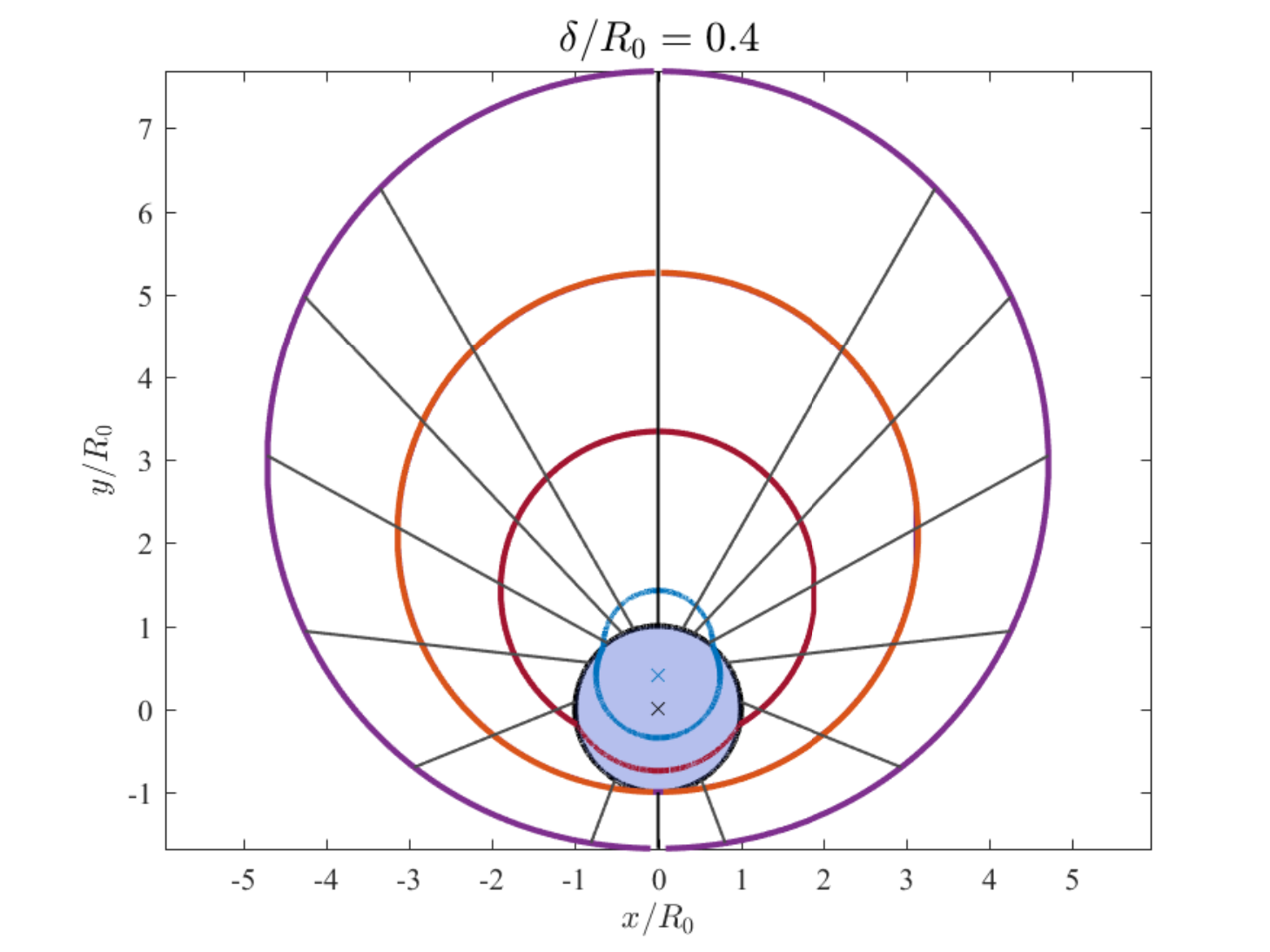}
    \caption{An off-center explosion within a uniform density sphere. The explosion is offset at a distance $\delta = 0.4 \, R_0$ from the center of the sphere.
    \textit{Colored contours} show the position of the shock front inside the sphere, and the extent of the outermost ejecta outside the sphere, at different times, before and after the shock broke out of the entire sphere surface. \textit{Black lines} are the straight trajectories of the ejecta.}
    \label{fig:OffsetExplosion04}
\end{figure}

\begin{figure}[!ht]
    \includegraphics[width=1.1\columnwidth]{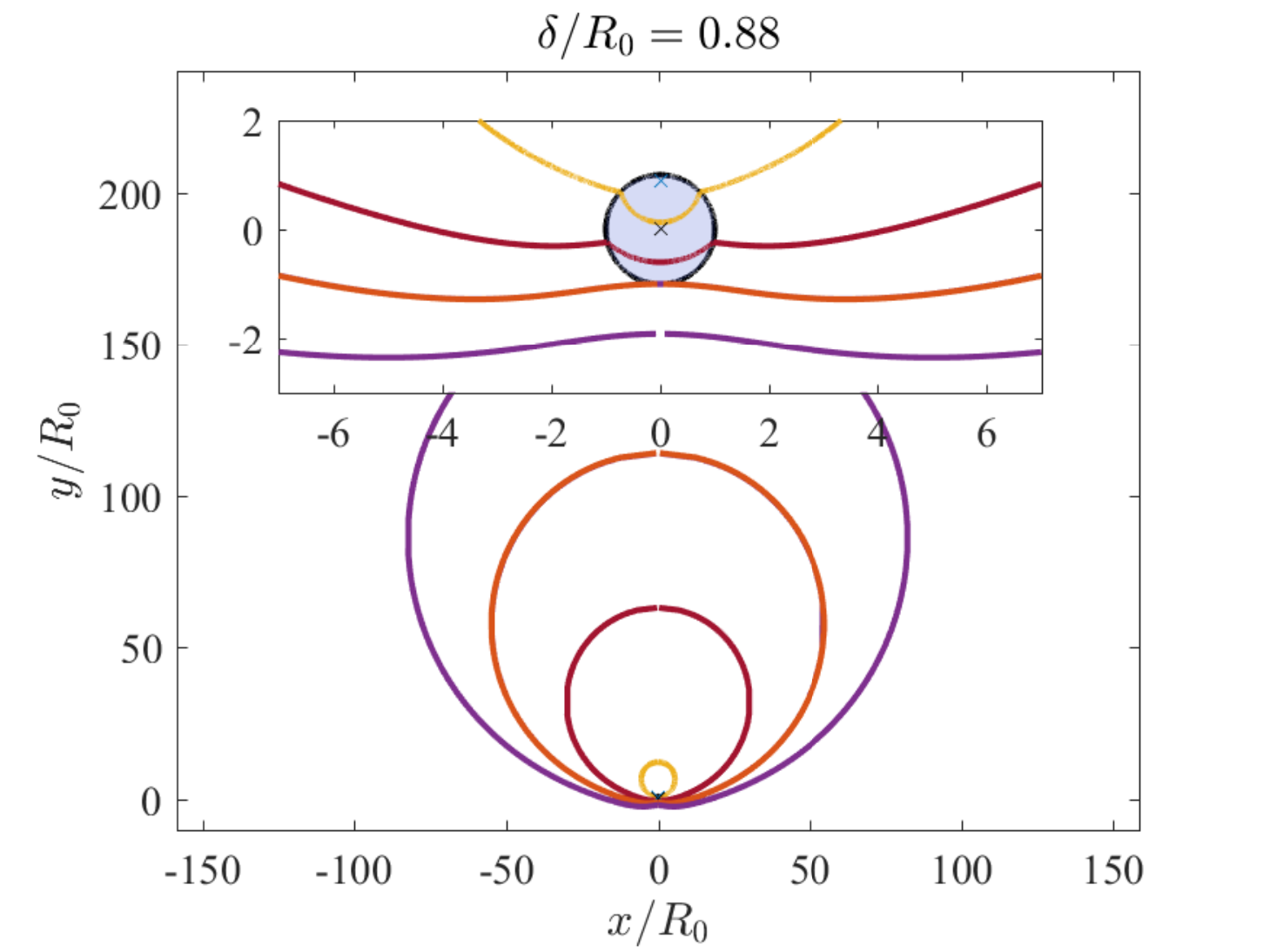}
    \caption{Same as figure \ref{fig:OffsetExplosion04}, for $\delta = 0.88 \, R_0$. This value is approaching the limit on $\delta$, for which our solution is valid, assuming $\gamma = 5/3$ (equation \ref{eq:delta_max}). The inset shows the ejecta envelope close to the sphere, where the slight concavity of the envelope is apparent.}
    \label{fig:OffsetExplosion088}
\end{figure}

Assuming $\delta < \delta_{max}$, our solution for a planar shock wave encountering a planar surface (section \ref{sec:analytical_solution}) can again be applied to calculate the propagation of ejecta from such an explosion. Repeating the calculation described in section \ref{subsec:surface_explosion}, we obtain the ejecta's envelope at any time. Figures \ref{fig:OffsetExplosion04} and \ref{fig:OffsetExplosion088} demonstrate the extent of the ejecta envelope at different times, for an explosion with $\delta/R_0 = 0.4$ and $\delta/R_0 = 0.88$. Remarkably, for large values of $\delta$, close to the limit $\delta_{max}$, the ejecta's envelope is concave near the antipodal point, as illustrated in the inset of figure \ref{fig:OffsetExplosion088}.

\section{Summary} \label{sec:discussion}
We discussed the hydrodynamics of an oblique shock breakout from a uniform density medium. We derived an analytic steady-state solution for an arbitrary shock-boundary intersection angle $\beta$. The steady-state frame follows the shock-boundary intersection, travelling at the pattern velocity $v_0$. The lack of a length-scale near the shock-boundary intersection implies that the flow fields depend on the azimuthal angle $\theta$ alone, reducing this two-dimensional problem to a a set of coupled ordinary differential equations.

The flow is separated to a uniform density region, where matter flows along straight streamlines, just past the shock, up to an angle $\theta_\star$. The flow then rarefies as a Prandtl-Meyer expansion fan, terminating at an angle $\theta_f$, at which cold matter flows purely radially with velocity $v_0$. The steady-state solution exists \textcolor{black}{for a range of obliquity angles, up to some maximum angle, $\beta_{max}$}. \textcolor{black}{Beyond this critical angle}, sound waves emitted from the expanded flow can overtake the shock front.

At the limit $\beta \ll 1$, our solution reproduces the one-dimensional flow obtained when a planar shock breaks out to vacuum. The pattern velocity $v_0$ diverges as $\beta$ tends to zero. Yet, only the normal velocity component (in the $y$ direction) remains after transforming from the steady-state frame back to the lab frame, resulting in a \textcolor{black}{one-dimensional} flow, normal to the surface, with vanishing lateral motion.

Applying these results, we considered a few scenarios in which our analytical solution can describe the flow close to the breakout point. The first is the breakout of \textcolor{black}{a} bow shock formed in a cold medium. A small object travelling within a uniform medium produces a parabolic shock front. If the medium terminates at a flat boundary, the shock will breakout to the vacuum, producing a spray of ejecta. Using our analytical result, we find that the outermost ejecta follows a straight line, forming an angle that scales as $R/d$, where $R$ is the size of the object, and $d$ is the separation of the object from the boundary.

As a second application, we consider a strong point explosion occurring near the surface of a uniform medium. This scenario could be applicable, for example, to strong underwater explosions, or to detonations under the surface of an asteroid. A spherical shock wave initially expands as a Sedov-Taylor explosion, until reaching the surface at time $t_0$. The shock breaks out at an increasingly oblique angle. Using our steady-state solution, we calculate the expansion of the outermost ejecta after the initial breakout. The intersection angle $\beta$ increases to $\beta_{max}$ at time $t_{max} \approx 7.5 \, t_0$ (for $\gamma = 5/3$). Up to time $t_{max}$ the shock continues to propagate as a Sedov-Taylor explosion within the uniform medium, oblivious to the presence of vacuum beyond the boundary. At $t_{max}$, sound waves originating from the expanded flow begin to catch up with the shock front, making our steady-state solution invalid at later times. Nonetheless, the outermost ejecta propagates ballistically and can thus still be followed at later times, $t>t_{max}$, even when the propagation of the shock inside the medium is altered. 
\textcolor{black}{Note that gravity was neglected in our analysis, unlike most works on underwater explosions (see \cite{Holt:1977}). This approximation is valid if $g$, the surface gravity, is much smaller than $\sqrt{R_0/t_0^2}$. Under this condition, gravity begins to shape the explosion's ejecta at late times, longer than $R/g \, t_0$, much later than $t_{max}$}.

Finally, we consider an off-center explosion within a sphere of uniform density. The shock wave emerges obliquely at the surface, and we use our solution to follow the evolution of the ejecta at any time. Remarkably, our analytical solution is applicable up to large offsets from the center of the sphere, with $\delta_{max}/R_0 \approx 0.89$ ($\gamma=5/3$) where $\delta$ is the explosion's offset from the center, and $R_0$ is the sphere's radius.

\begin{acknowledgments}
IL thanks support from the Adams fellowship. RS is supported by an ISF and an iCore grant.
\end{acknowledgments}

\section*{Appendix - Derivation of the flow equations} \label{sec:appendix}

\color{black}
An inviscid, adiabatic, compressible flow is be described by the flow equations in cylindrical coordinates as follows
\begin{equation} \label{eq:continuity_eq_full}
    \frac{\partial \rho}{\partial t} + \frac{1}{r} \frac{\partial(\rho r v_r)}{\partial r} + \frac{1}{r} \frac{\partial (\rho v_\theta)}{\partial \theta} + \frac{\partial (\rho v_z)}{\partial z} = 0 \,,
\end{equation}
\begin{equation} \label{eq:momemtum_r_eq_full}
    \rho \left( \frac{\partial v_r}{\partial t} + v_r \frac{\partial v_r}{\partial r} + \frac{v_\theta}{r} \frac{\partial v_r}{\partial \theta} + v_z \frac{\partial v_r}{\partial z} - \frac{v_\theta^2}{r} \right) = -\frac{\partial p}{\partial r} \,,
\end{equation}
\begin{equation} \label{eq:momemtum_theta_eq_full}
    \rho \left( \frac{\partial v_\theta}{\partial t} + v_r \frac{\partial v_\theta}{\partial r} + \frac{v_\theta}{r} \frac{\partial v_\theta}{\partial \theta} + v_z \frac{\partial v_\theta}{\partial z} + \frac{v_r v_\theta}{r} \right) = -\frac{1}{r} \frac{\partial p}{\partial \theta} \,,
\end{equation}
\begin{equation}
    \rho \left( \frac{\partial v_z}{\partial t} + v_r \frac{\partial v_z}{\partial r} + \frac{v_\theta}{r} \frac{\partial v_z}{\partial \theta} + v_z \frac{\partial v_z}{\partial z} \right) = -\frac{\partial p}{\partial z} \,,
\end{equation}
representing the continuity and momentum equations. Since the flow in consideration is steady, $\partial / \partial t = 0$. The flow is two-dimensional, hence there is no $z$ dependence, $v_z = 0$ and $\partial / \partial z = 0$. Finally, the flow fields $\rho$, $p$, $v_r$ and $v_\theta$ are independent of $r$. The continuity equation (\ref{eq:continuity_eq_full}) than reduces to equation \ref{eq:continuity_uniform_density}, the radial momentum equation (\ref{eq:momemtum_r_eq_full}) reduces to \ref{eq:momentum_r_uniform_density}, and the tangential momentum equation (\ref{eq:momemtum_theta_eq_full}) reduces to 
\begin{equation} \label{eq:momentum_theta_intermediate}
    v_\theta \frac{\partial v_\theta}{\partial \theta} + v_r v_\theta = -\frac{1}{\rho} \frac{\partial p}{\partial \theta} \,.
\end{equation}

Finally, assuming a polytropic equation of state we replace the pressure in equation \ref{eq:momentum_theta_intermediate} by the speed of sound, using $c_s^2 = \gamma p / \rho$, to obtain equation \ref{eq:momentum_theta_uniform_density}.

\color{black}

\nocite{*}
\bibliography{obliquebreakout}

\end{document}